\documentclass[useAMS,usenatbib]{mn2e}
\usepackage{epsfig,amsmath,natbib}
\def\be{\begin{equation}} 
\def\ee{\end{equation}}

\def\gsim{\lower.5ex\hbox{\gtsima}} 
\def\lsim{\lower.5ex\hbox{\ltsima}} \def\gtsima{$\; \buildrel > \over 
\sim \;$} \def\ltsima{$\; \buildrel < \over \sim \;$} \def\prosima{$\; 
\buildrel \propto \over \sim \;$} \def\gsim{\lower.5ex\hbox{\gtsima}} 
\def\lsim{\lower.5ex\hbox{\ltsima}} 
\def\simgt{\lower.5ex\hbox{\gtsima}} 
\def\simlt{\lower.5ex\hbox{\ltsima}} 
\def\simpr{\lower.5ex\hbox{\prosima}} \def\la{\lsim} \def\ga{\gsim} 
  
\def\gtsima{$\; \buildrel > \over \sim \;$} 
\def\ltsima{$\; \buildrel < \over \sim \;$} 
\def\gsim{\lower.5ex\hbox{\gtsima}} 
\def\lsim{\lower.5ex\hbox{\ltsima}} 
\def\simgt{\lower.5ex\hbox{\gtsima}} 
\def\simlt{\lower.5ex\hbox{\ltsima}} 
\def\simpr{\lower.5ex\hbox{\prosima}} 
\def\la{\lsim} 
\def\ga{\gsim} 
\def\Zcr{Z_{\rm cr}}

\def\Msun{M_{\odot}} 
\def\Zsun{Z_{\odot}} 
\def\Lsun{L_{\odot}}

\def\E3{{\cal E}_{\rm g}^{III}}

\title{Carbon-enhanced metal-poor stars in dwarf galaxies}
\author[Salvadori, Sk\'ulad\'ottir \& Tolstoy]
{Stefania Salvadori$^{1,\star}$, \'Asa Sk\'ulad\'ottir$^{1}$ \& Eline Tolstoy$^{1}$\\
$^1$Kapteyn Astronomical Institute, University of Groningen, Landleven 12, 
9747 AD Groningen, The Netherlands\\
$\star$VENI Fellow}
\begin{document} 
\date{} 
\pagerange{\pageref{firstpage}--\pageref{lastpage}} \pubyear{} 
\maketitle 
\label{firstpage} 
\begin{abstract}
%%%%%%%%%%%%%%%%%%%%%%%%%%%%%%%%%%%%%%%%%%%%%%%%%%%%%%%%%%%%%%%%%%%%%%%%%%%%%%%%
We investigate the frequency and origin of carbon-enhanced metal-poor 
(CEMP) stars in Local Group dwarf galaxies by means of a statistical, 
data-calibrated cosmological model for the hierarchical build-up of the 
Milky Way and its dwarf satellites. The model self-consistently explains 
the variation with dwarf galaxy luminosity of the observed: i) frequency 
and [Fe/H] range of CEMP stars; ii) metallicity distribution functions; 
iii) star formation histories. We show that if primordial faint supernovae 
dominated the early metal enrichment, then CEMP-no stars enriched by the 
first stellar generations should be present in {\it all} dwarf galaxies, 
with similar number of stars and CEMP fractions at [Fe/H]$< -4$. We 
demonstrate that the probability to observe a star that is carbon-enhanced 
within a given [Fe/H] range strongly depends on the luminosity of the dwarf 
galaxy and, on average, it is an order of magnitude lower in ``classical'' 
Sculptor-like dSph galaxies ($P\leq 0.02$) than in the least luminous 
ultra-faint dwarfs ($P \approx 0.1$). In addition, we explain why it may 
be easier to find CEMP-no stars at [Fe/H]$\approx -2$ in classical dSph 
galaxies than in ultra-faint dwarfs. These are consequences of the 
dramatic variation in the fraction of stars at [Fe/H]$<-3$ with galaxy 
luminosity: $\geq 40\%$ for galaxies with $L<10^5\Lsun$, and $\leq 0.2\%$ 
for $L>10^{7}\Lsun$. We present model predictions for the low Fe-tail and 
CEMP fraction of stars in dwarf galaxies, with particular emphasis on the 
Sculptor dSph, that can be used to shed light on the properties of the first stars.
\end{abstract}
%%%%%%%%%%%%%%%%%%%%%%%%%%%%%%%%%%%%%%%%%%%%%%%%%%%%%%%%%%%%%%%%%%%%%%%%%%%%%%%
\section{Introduction}
%%%%%%%%%%%%%%%%%%%%%%%%%%%%%%%%%%%%%%%%%%%%%%%%%%%%%%%%%%%%%%%%%%%%%%%%%%%%%%%
Understanding the nature of the first stars is one of the key questions
of modern Cosmology.
Despite extensive observational searches, truly pristine
stars have so far escaped detection. Currently, the most metal-deficient star 
known has a total metallicity of $Z\approx6.9\times10^{-7}$ \citep{caffau2011}. 
The persistent lack of more metal-poor stars seems to confirm the idea that 
primordial stars were all more massive than $\approx 1\Msun$ \citep[e.g.
][]{mckee08,hosokawa11,hirano14}, and that their formation was possibly 
quenched at early times \citep[e.g.][]{salvadori07,pallottini14}. 
However, the chemical signatures of these extinct stellar generations 
could be retained in the photospheres of ancient ($>12$~Gyr) low-mass 
second-generation stars, which formed in pre-enriched environments, 
$Z>Z_{cr}=10^{-5\pm1}\Zsun$, where metals and dust grains dispersed by 
the first stars enabled efficient gas cooling and fragmentation
\citep[e.g.][]{schneider02}. These stellar fossils should be observable 
in the oldest stellar components of our Galaxy and in its ancient 
dwarf satellites.

High- and medium-resolution spectroscopic studies of Galactic halo stars 
have revealed the existence of a population of carbon-rich stars 
\citep[e.g.][]{beers05,aoki07,yong13,norris13,lee13}. These objects are usually 
defined to have carbon-to-iron ratio [C/Fe]$>0.7$ \citep{aoki07}, and they 
can be divided into two main populations: carbon-rich stars that exhibit an 
excess in heavy elements formed by slow (or rapid) neutron capture processes, 
CEMP-s (CEMP-r) stars, and carbon-rich stars that do not exhibit such an
excess, CEMP-no stars. The available data are consistent with the idea that 
CEMP-s stars belong to binary systems \citep{lucatello05,starkenburg14}, 
and have acquired their carbon-excess from an Asymptotic Giant Branch (AGB) 
companion \citep[e.g.][]{bisterzo12,placco13,abate15}. Thus, the chemical 
abundances measured in these stars are not representative of the interstellar 
medium (ISM) out of which they formed.
On the other hand, CEMP-no stars are {\it not} preferentially associated 
to binary systems \citep{hansen13,cohen13,norris13,starkenburg14}. Hence 
there is no observational evidence supporting the idea of mass transfer 
as the origin of their chemical abundances, which was suggested by some 
authors \citep[e.g.][]{suda04,komiya07}.
Furthermore, both their {\it frequency} and {\it carbon-excess} increase with
decreasing [Fe/H] \citep[e.g.][]{lucatello05,lee13,norris13}, and 8 out of the 
9 halo stars discovered at [Fe/H]$<-4.5$ are CEMP-no stars \citep{christlieb02,frebel05,norris07,keller14,hansen15,bonifacio15,allende-prieto15,frebel15}.
These findings favor the idea that CEMP-no stars are a peculiar stellar
population and that their chemical abundances likely reflect their birth 
environment.

The unusual chemical compositions of the most iron-poor and carbon-rich 
stars can be successfully matched by models of {\it primordial faint SN} 
that experienced mixing and fallback, hence releasing small amounts of iron 
and large amounts of carbon and other light elements \citep[e.g.][]{
umeda03,iwamoto05,joggerst09,marassi14,tominaga14}. 
A reatively good agreement with observations is also obtained by models of
zero- (or very low-) metallicity massive ``spinstars'', which experience 
mixing and mass loss because of their very high-rotational velocities 
\citep[e.g.][]{meynet06,meynet10,meynet15}. Recently, chemical evolution 
studies have further support the idea of the link between primordial {\it faint} 
SN and CEMP-no stars, showing that the observed fraction of carbon-enhanced 
to carbon-normal stars at [Fe/H]$<-3$ is successfully reproduced {\it if 
faint SN dominated the early metal-enrichment} \citep{debennassuti14,cooke14}. 
Thus, we can work under this simple hypothesis to predict the frequency of 
pristine carbon-enhanced stars in {\it ancient} and {\it metal-poor} dwarf
galaxies.

Carbon-enhanced metal-poor stars have been found in a significant fraction 
in the faintest satellites of the Milky Way, the so-called ultra-faint 
dwarf galaxies, with total luminosities $L\leq 10^5\Lsun$ 
\citep[e.g.][]{norris10,frebel10a,lai11,gilmore13,frebel14}. 
Different groups have 
proposed these galaxies to be the living relics of the first star-forming 
minihaloes, which formed in the Milky Way environment prior the end of
reionization
\citep[e.g.][]{salvadori09,bovill09,munoz09,bovill11,salvadori12,salvadori14}. These theoretical predictions are consistent with recent observations of star formation histories in ultra-faint dwarf galaxies 
\citep[e.g.][]{dallora12,okamoto12,brown12,brown14}. 
In particular, by interpreting the observed Fe-Luminosity relation and
metallicity distribution function (MDF) of dwarf galaxies in a cosmological context, \cite{salvadori09} predicted these ancient ultra-faint dwarf 
galaxies to be the best objects to look for the chemical imprints of 
the first stellar generations. This picture is supported by the recent 
discovery of several CEMP-no stars at [Fe/H]$<-3$ in Segue 1 \citep{frebel14},
which is one of the faintest ultra-faint dwarfs.

However, carbon-enhanced metal-poor stars seem to be rare in the more 
luminous ``classical'' dwarf spheroidal (dSph) galaxies, $L > 10^5\Lsun$,
where measurements are available for larger stellar samples. The deficiency 
of CEMP(-no) stars is especially mysterious in the Sculptor dSph galaxy. 
The observed Color-Magnitude-Diagram (CMD) and MDF of Sculptor are consistent 
with this galaxy being dominated by ancient stars, $>10$~Gyr old 
\citep{deboer12}. However, no CEMP stars have been found among the ten
carefully studied stars at [Fe/H]$<-3$ \citep{frebel10b,tafelmeyer10,
starkenburg13,jablonka15,simon15}. 
Only recently, the first CEMP{\it-no} star has been discovered at [Fe/H]$\approx -2$ 
\citep{skuladottir15}. The observed chemical abundance pattern of 
this star is consistent with the idea that the carbon excess is the 
result of a {\it pristine} population of faint SN polluting its 
birth-environment. However, the lack of CEMP-no stars at lower [Fe/H] 
is not. The important question is, are carbon-enhanced {\it metal-poor} 
stars missing in classical dSph galaxies, or are they hidden?

We argue they may be hidden. To sustain this thesis, in this paper we use 
a statistical, data-calibrated cosmological model for the formation of the 
Milky Way and its dwarf satellites, to predict the frequency of CEMP(-no) 
stars in Sculptor and other Local Group dwarf galaxies. We present a simple, 
global scenario that self-consistently explains the variation with galaxy 
luminosity of the observed: i) carbon-enhanced to carbon-normal star ratios; 
ii) metallicity distribution functions; and iii) star formation histories. 
We link the properties of Local Group dwarf galaxies with first stellar 
generations and early galaxy formation processes, and present model 
predictions aimed at identifying the hidden CEMP-no stars in the classical 
dSph galaxy Sculptor. 

For consistency with previous works \citep{salvadori09,salvadori12,salvadori14}, 
along with Via Lactea II, Aquarius, and CLUES simulations \cite[e.g.][]{madau08,wang12,navarro15}, we adopt a Lambda Cold Dark Matter ($\Lambda$CDM) cosmology with 
${\Omega}_m=0.24$, ${\Omega}_{\Lambda}=1-{\Omega}_m=0.76$, ${\Omega}_b=0.04$, 
and $H_0=73$km/s/Mpc. Furthermore, in the overall paper we assume the solar 
abundance values by \cite{asplund09}.
%%%%%%%%%%%%%%%%%%%%%%%%%%%%%%%%%%%%%%%%%%%%%%%%%%%%%%%%%%%%%%%%%%%%%%%%%%%%%
\begin{figure}   
\begin{centering}
  \includegraphics[width=1.05\columnwidth]{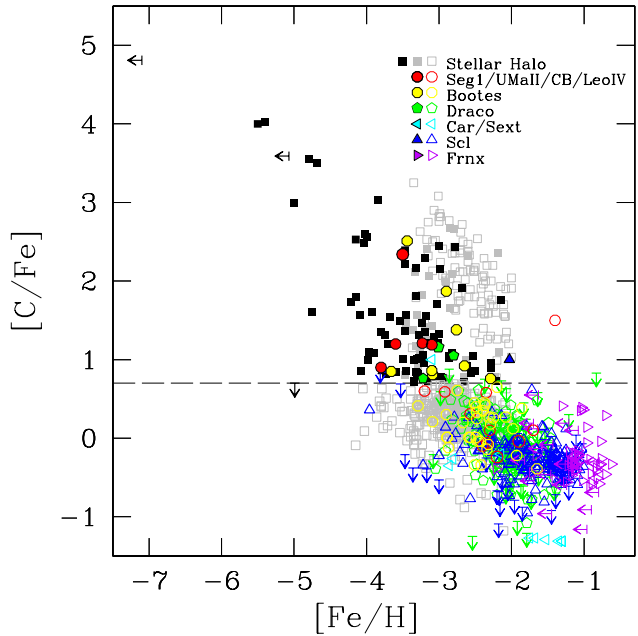}
    \caption{Compilation of stars with measured [C/Fe] and [Fe/H] 
    in {\it the stellar halo} (squares), {\it ultra-faint} dwarf 
    galaxies (circles, hexagones, pentagones) and {\it classical} 
    dSph galaxies (triangles).     
    [C/Fe] measurements are corrected to account for internal mixing 
    processes (see text). {\it CEMP-no stars} are shown as filled 
    symbols, upper limits with arrows. Stars with [C/Fe]$>0.7$ and 
    open symbols are {\it CEMP-s/r stars}. Filled gray squares are 
    CEMP stars with no available measurements of r- or s-process 
    elements. Colors/symbols identify stars in dwarf galaxies with 
    increasing total luminosity: from red to blue (see labels and text).
    References: 
    {\it Halo stars:} \cite{placco14,christlieb02,frebel05,norris07,
    caffau2011,keller14,hansen15,bonifacio15,frebel15}.  
    {\it Segue 1:} \cite{norris10,frebel14}. 
    {\it Ursa Major II} and {\it Coma Berenice:} \cite{frebel10a}.
    {\it Leo IV:} \cite{simon10}. 
    {\it Bootes:} \cite{lai11,norris10,gilmore13}. 
    {\it Draco:} \cite{cohen09,shetrone13,kirby15}. 
    {\it Sextans:} \cite{honda11}.
    {\it Carina:} \cite{venn12}. 
    {\it Sculptor:} \citep{frebel10b,tafelmeyer10,
    starkenburg13,simon15,kirby15,jablonka15,skuladottir15}.
    {\it Fornax:} \citep{kirby15}.} 
    \label{Fig:data}
\end{centering}
\end{figure}
%%%%%%%%%%%%%%%%%%%%%%%%%%%%%%%%%%%%%%%%%%%%%%%%%%%%%%%%%%%%%%%%%%%%%%%%%%%%
\section{Observations of carbon-rich stars}    
%%%%%%%%%%%%%%%%%%%%%%%%%%%%%%%%%%%%%%%%%%%%%%%%%%%%%%%%%%%%%%%%%%%%%%%%%%%%%
Current observations of CEMP stars, [C/Fe]$>0.7$, are shown in Fig.1, 
where we present a sample of stars from the literature with measured 
carbon-to-iron ratio, [C/Fe], and iron-abundance, [Fe/H], in both the 
Galactic halo and nearby dwarf galaxies. Most of these data are 
based on one dimensional local thermodynamic equilibrium (1D/LTE) 
stellar model analysis.

Measurements for {\it halo stars} are from \cite{placco14}, who selected 
among all available data (see references therein) a sample of 505 stars 
with [Fe/H]$<-2$ and [C/Fe] measurements. This sample includes dwarf and
giant stars ($0<log\,g < 5$). \cite{placco14} corrected [C/Fe] to account 
for the depletion of surface carbon abundance, which is expected to occur 
on the upper Red Giant Branch (RGB), $log\,g < 2$. 
The correction depends on several observed quantities: $log\,g$, [C/Fe], 
[Fe/H] and also, to a lesser extent, on [N/Fe]. Hence, we should keep in 
mind that the corrected values presented in Fig.~1 have intrinsic errors, 
which are estimated by the authors to be always within the $2\sigma$ 
uncertainties of the [C/Fe] measurements, i.e. $\pm 0.3$ dex.

In more distant dwarf galaxies only the brighter RGB stars are 
typically observed. Hence, in our selection of dwarf galaxy stars, 
if the internal mixing was not already accounted for by the authors, 
we used the online tool by \cite{placco14} to self-consistently 
correct the [C/Fe] measurements. 
When not available from observations we simply assumed [N/Fe]$=0.0$, 
in agreement with \cite{placco14}. The data shown in Fig.~1 represents 
the largest sample of [C/Fe] measurements that have been homogeneously 
corrected to account for internal mixing processes.\\

From Fig.~1 we can note that stars at [Fe/H]$\leq -4.5$ have only 
been found in the Galactic halo. Moreover, we can see that the 
frequency of CEMP-no stars among these ``hyper-iron-poor stars'' 
is extremely high: 8 out of 9 stars at [Fe/H]$\leq-4.5$ are CEMP-no 
stars. These objects have peculiar chemical abundance patterns 
consistent with the yields predicted for primordial faint SN 
\citep[e.g.][]{iwamoto05}. However, we also note that these extreme 
stars are part of a common, global trend, which involves stars in 
both the Galactic halo and dwarf galaxies. We see that on average 
[C/Fe] is higher in stars with lower [Fe/H], and it declines with 
[Fe/H]. The same trend affects the incidence of CEMP-no stars with 
respect to the overall stellar population. At [Fe/H]$\leq-3.0$, the 
fraction of CEMP-no stars in the Galactic halo, ($\approx 43\%$,
Placco et al. 2014), is consistent with the {\it overall} fraction 
of CEMP-no stars in dwarf galaxies, where 24 stars at [Fe/H]$\leq-3$ have 
been observed, and 10 of them are found to be CEMP-no stars. This gives 
$F_{\tt CEMP}(<-3)\approx 42\%$. However, when we consider individual 
dwarf galaxies, the fractions are highly variable.

The carbon measurements in the least luminous ultra-faint dwarf galaxies, 
$log(L/\Lsun) < 4.0$ (shown in Fig.~1 with red circles), are from 
high-resolution spectroscopic studies (see caption for references). Although 
less than 20 stars have been observed in these 4 systems, the overall 
frequency of CEMP-no stars is very high, and at [Fe/H]$\leq-3$, 5 out 
of 6 stars are CEMP-no stars, $F_{\tt CEMP}(\leq-3)\approx 83\%$. 
{With the only exception of seven stars in Bootes, one of which 
is a CEMP-no at [Fe/H]$=-3.5$ \citep{gilmore13}, only low resolution 
spectroscopic studies are available for dwarf galaxies with luminosities 
between ultra-faint and classical dSph galaxies, e.g. {\it Bootes}, 
$log (L/\Lsun)=4.5\pm 0.1$, and {\it Draco}, $log (L/\Lsun)=5.45\pm0.10$. 
Thus, slow and rapid n-capture elements have not been measured in these stars, 
preventing us from distinguishing between CEMP-no and CEMP-r/s stars. 
As discussed in Sec.~1, however, there is strong observational 
evidence that CEMP-no stars should be the dominant CEMP population 
at [Fe/H]$<-3$ \citep[e.g.][]{norris13}. In these dwarf galaxies
the fraction of CEMP(-no) stars at [Fe/H]$\leq-3$ is high: 4 out of 6 stars 
in Bootes ($\approx 66\%$), and $2$ out of $5$ stars in Draco ($\approx 
40\%$). The classical dSph galaxies {\it Carina} and {\it Sextans}, 
$log (L/\Lsun)\approx 5.6$, have been followed up at high-resolution,
but only stars at [Fe/H]$\geq -3$ have carbon measurements (see Fig.~1). 

In the {\it Sculptor} dSph galaxy, $log (L/\Lsun)=6.34\pm 0.16$, 
many carbon measurements are available from both low- 
\citep{kirby15}, and high-resolution spectroscopic studies
\citep{frebel10b,tafelmeyer10,starkenburg13,simon15,skuladottir15}. 
In total, 10 stars have been found at [Fe/H]$\leq-3$. 
After correcting for internal mixing, we find that the star 
Scl$\_11\_1\_4296$ observed by \cite{simon15} at [Fe/H]$=-3.77$ 
can be possibly identified as a CEMP-no star, [C/Fe]$=0.77
\pm 0.34$. This was not claimed by the authors, who used the
luminosity dependent criteria developed by \cite{aoki07} to
identify CEMP stars\footnote{There are small differences 
between the \cite{aoki07} criteria and the carbon correction by
\cite{placco14}.}.However, the carbon-excess is small and the errors big, making this 
classification very uncertain. Thus, the only reliable CEMP-no 
star discovered in Sculptor, [C/Fe]$=1.01\pm 0.1$, has an unusually 
high iron-abundance, [Fe/H]$\approx -2$ \citep{skuladottir15}, 
which makes this star stand out with respect to the general trend 
in dwarf galaxies (Fig.~1)}. No CEMP-no stars have been found in 
Fornax, $log (L/\Lsun)\approx 7.25\pm 0.11$, although carbon 
measurements are only available 
for stars at [Fe/H]$>-2$. For other ultra-faint dwarfs 
(e.g. Segue 2, Willmann, Hercules) or classical dSph galaxies 
(e.g. LeoT, LeoI, LeoII, Ursa Minor) carbon measurements are not 
yet available. 

In conclusion, while $F_{\tt CEMP}(\leq-3)$ in the Galactic halo is 
consistent with the {\it overall} fraction of CEMP-no stars at 
[Fe/H]$\leq-3$ in dwarf galaxies, the individual Milky Way companions
show that $F_{\tt CEMP}(\leq-3)$ strongly decreases when the luminosity 
of the galaxies increases. Does the fraction of CEMP-no stars depend on 
galaxy luminosity? And why does the only CEMP-no star observed in a 
more luminous dSph galaxy have an unusually high [Fe/H]$\approx -2$? 
Are these observational findings reconcilable with the idea that the 
birth environment of CEMP-no stars was polluted by primordial faint SN, 
and thus consistent with what we see in the Galactic halo?
%%%%%%%%%%%%%%%%%%%%%%%%%%%%%%%%%%%%%%%%%%%%%%%%%%%%%%%%%%%%%%%%%%%%
\section{Cosmological merger-tree model}    
%%%%%%%%%%%%%%%%%%%%%%%%%%%%%%%%%%%%%%%%%%%%%%%%%%%%%%%%%%%%%%%%%%%%
We use the data-constrained cosmological code GAMETE \cite[GAlaxy 
MErger Tree and Evolution,][]{salvadori07,salvadori09,salvadori12,
salvadori14,debennassuti14} to link the properties of ancient stars 
in the Local Group with the early star formation and metal-enrichment 
processes. GAMETE describes the formation of the Galaxy and its dwarf 
satellites in a $\Lambda$CDM framework, self-consistently accounting 
for the key physical processes driving the formation and evolution 
of high-redshift dwarf galaxies: 
(i) the transition from {\it massive}, and hence short-lived 
Population III (Pop~III) stars, to {\it normal} Population II 
(Pop~II) stars; 
(ii) the gradual quenching of star formation in dwarf galaxies with increasing total masses due to the enhanced 
photo-dissociating and photo-ionizing radiation; (iii) the steady 
metal-enrichment of the Milky Way environment, or {\it Galactic Medium}, 
due to supernovae-driven outflows from star forming haloes. 

Our model is a statistical tool that enables us to study the most 
likely assembly and metal-enrichment histories of Local Group galaxies. 
The star formation and chemical evolution of present-day galaxies is traced 
across cosmic time by exploiting a sample of merging histories of a Milky 
Way-size dark matter halo ($M_{MW}=10^{12}\Msun$), reconstructed via a Monte 
Carlo algorithm based on the Extended Press-Schechter formalism 
\citep[see][for more details]{salvadori07,debennassuti14}. 
Newly collapsed haloes are assumed to have a gas-to-dark matter mass 
ratio equal to the baryonic cosmic fraction, $M_g/M=\Omega_b/\Omega_M$
and a chemical composition equal to the Milky Way environment at their 
formation epoch. Hence they are all primordial, $Z=0$, before the onset 
of supernovae explosions. The main and new features of the model are 
summarized below, including the underlying key assumptions relevant for 
this work. We refer the reader to previous papers for more details.\\

{\bf Star formation (SF).} 
The SF is traced along the merger trees by adopting physically 
motivated hypotheses. 
\begin{itemize}
\item There exists a minimum halo mass to form stars, $M_{sf}(z)$, 
whose evolution accounts for the suppression of SF in progressively 
{\it more massive} objects due to the increasing photo-dissociating 
and photo-ionizing radiation \cite[][]{salvadori09,salvadori12}. 
When the Milky Way environment is fully reionized, $z_{rei}\approx 6$ 
\citep{salvadori14}, we assume that gas accretion is suppressed in 
haloes with virial temperatures $T_{vir}\simlt 2\times 10^4$~K.
\item The SF rate, $\psi=\epsilon_* M_g/t_{ff}$, which is regulated 
by the SF efficiency, $\epsilon_*$, depends on the free-fall time, 
$t_{ff}(z)$, and mass of cold gas in each galaxy, whose gradual accretion 
is described by a numerically calibrated infall rate \citep{salvadori08}.
\item In {\it minihaloes} with $T_{vir}\simlt10^4$~K, the SF efficiency 
is assumed to be reduced as $\epsilon_{H_2}=2{\epsilon_*}[1+\big(T_{vir}/{2\times10^4K}\big)^{-3}]^{-1}$ to account for the ineffective 
cooling by molecular hydrogen, H$_2$ \citep{salvadori12}.
\item Pop~II stars with masses $m=[0.1-100]\Msun$ form according 
to a Larson Initial Mass Function, $\Psi(m)=m^{-2.35}e^{-0.35\Msun/m}$, 
if the gas metallicity exceeds the critical value, $\Zcr$, which sets 
the minimal conditions for the formation of the first low-mass stars.
This value can be either $Z_{cr}=(10^{-4}-10^{-3})\Zsun$, if gas 
fragmentation is driven by metal-line cooling \citep[e.g.][]{bromm01,frebel09},
or $Z_{cr}=(10^{-6}-10^{-4})\Zsun$, if it is due to thermal emission 
by collisionally excited dust grains \citep[e.g.][]{schneider02,omukai05}. 
Following the most recent findings we set $\Zcr=10^{-4.15}\Zsun$ 
\citep[e.g.][]{caffau2011,schneider12,debennassuti14}, and we explore the case
$\Zcr=10^{-6}\Zsun$ in Sec.~6.
\item Massive Pop~III stars form if $Z<Z_{cr}$. To work under the hypothesis 
that the early metal-enrichment is dominated by faint SN \citep[e.g.][]{cooke14,
debennassuti14}, we adopt the simplest prescription and assume that all Pop~III 
stars have a characteristic stellar mass of $25\Msun$ and evolve as {\it faint 
supernovae}, which experience mixing and fallback \citep[e.g.][]{salvadori12}. 
\end{itemize}

{\bf Chemical enrichment.}
The contribution to the chemical enrichment by different stellar populations 
is traced in the code by accounting for the mass and metallicity dependent 
stellar lifetimes, and by including all chemical elements from C to Zn 
\citep{salvadori12}. For Pop~III stars evolving as faint supernovae we 
adopt the yields by \cite{iwamoto05} for the case\footnote{The total amount
of Fe and C per stellar mass formed is equivalent to the integrated contribution
of faint SN with $m=(10-40)\Msun$, as their yields are rescaled with stellar
masses (e.g. Fig.2 by \cite{debennassuti14}) $25\Msun$ and kinetic explosion 
energy $E_{\tt PopIII}=0.7\times 10^{51}$erg}. Following \cite{woosley95} we 
assume that stars with 
$m > 40\Msun$ do not contribute to metal enrichment, and we adopt their 
yields systematically halved in iron \citep{timmes95} for massive $8\Msun
\leq m\leq 40\Msun$ stars that evolve as Type II supernovae (SNII), $\langle 
E_{\tt SN_{II}}\rangle=1.2\times 10^{51}$erg. For low and intermediate mass 
Asymptotic Giant Branch (AGB) stars, $m < 8\Msun$, we adopt the yields by 
\cite{hoek97} for metallicities $Z>10^{-3}$  and by \cite{meynet02} for 
$Z\leq10^{-5}$. All relevant equations describing the chemical enrichment 
of star-forming galaxies can be found in \cite{salvadori08}. 

The contribution of supernovae Type Ia (SNIa) has been included by 
adopting the yields and explosion energy ($E_{\tt SNIa}=1.3\times10^{51}$erg)
by \cite{iwamoto99}, and the bimodal delay time distribution 
observationally derived by \cite{mannucci06}. At each time-step and for
each star-forming halo of the merger tree we compute the rate of SNIa by
following \cite{matteucci06} and we fix the normalization constant, 
$A_{\tt SNIa}$, to reproduce the actual rate of SNIa in the Milky Way, $\approx(0.3/100)$yr$^{-1}$ \citep{cappellaro99}.

The chemical evolution of the gas is simultaneously traced in the 
star-forming haloes and in the surrounding {\it Galactic Medium}, 
or Milky Way environment, by including the effect of SN-driven outflows, 
which are regulated by the SN wind efficiency, ${\epsilon}_w$. During 
SN explosions, metals present in the interstellar medium (ISM) are ejected 
out of the galaxy at a rate $\dot{M}^{ej}_Z=Z\dot{M}^{ej}_g$, where 
$Z=M_Z/M_g$ is the metallicity of the gas, and $\dot{M}^{ej}_g$ the
gas ejection rate, which depends on the cumulative SN explosion energy, 
and on the binding energy of the host halo \citep[e.g.][]{salvadori08,debennassuti14}. Heavy elements dispersed into the Milky Way environment 
are assumed to be instantaneously mixed in this medium, so its total
metallicity, $Z_{GM}$, steadily increases across cosmic times.\\ 
%%%%%%%%%%%%%%%%%%%%%%%%%%%%%%%%%%%%%%%%%%%%%%%%%%%%%%%%%%%%%%%%%%%
\begin{figure}
  \includegraphics[height=0.76\columnwidth]{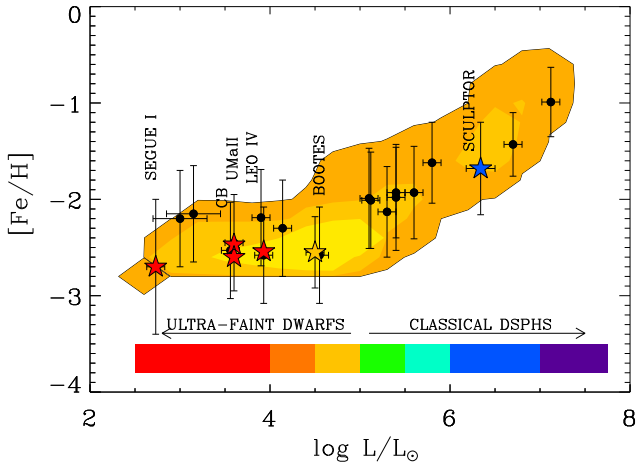}
   \caption{Predicted (contours) and observed (points) Fe-L
   relation for Milky Way dwarf galaxies. Contours identify 
   regions containing the $(68$,$95$,$99)\%$ of the dSph 
   candidates in 50 hierarchical merger histories. Observed 
   points are from \cite{kirby11}. Stars and labels 
   underline individual dwarf galaxies.}
    \label{Fig:FeL}
\end{figure}

%%%%%%%%%%%%%%%%%%%%%%%%%%%%%%%%%%%%%%%%%%%%%%%%%%%%%%%%%%%%%%%%%%%%%%%
{\bf Universality of the free parameters.}
%%%%%%%%%%%%%%%%%%%%%%%%%%%%%%%%%%%%%%%%%%%%%%%%%%%%%%%%%%%%%%%%%%%%%%%
The free parameters of the model are the star formation 
and SN-wind efficiency ($\epsilon_*$, $\epsilon_w$), the critical 
metallicity ($Z_{cr}$), and the fraction of stars that can give 
rise to SN Ia ($A_{\tt SNIa}$). These are fixed to simultaneously reproduce 
the global properties of the Milky Way at $z=0$, along with the 
MDF of Galactic halo stars \citep[e.g.][]{salvadori07,debennassuti14}. 
These free parameters are assumed to be {\it the same} for {\it all} 
the haloes 
of the merger tree and in {\it all} hierarchical merger histories of
the Milky Way. Hence, the properties of different dwarf galaxies we will 
present in the following, {\it are not the result of a fine tuning of the 
free parameters, but a consequence of the cosmological context in 
which we study the formation and evolution of these small systems.}\\

%%%%%%%%%%%%%%%%%%%%%%%%%%%%%%%%%%%%%%%%%%%%%%%%%%%%%%%%%%%%%%%%%%%%%%%
{\bf Selection and properties of satellite candidates.} 
Galaxies that can survive as satellites of the Milky Way are selected 
among star-forming haloes of the merger trees that at any given 
redshift have dark matter masses that correspond to low-sigma 
density fluctuations, $M< M_{2.5\sigma}(z)$. This assumption is 
supported by N-body simulations \citep[e.g.][]{diemand05}. 
After selection, its subsequent evolution is followed in isolation 
down to $z=0$ \citep{salvadori09}. 

In Fig.~2 we show the predicted Fe-Luminosity relation for all 
selected candidates in 50 assembly histories of the Milky Way. 
We assume $M/L=1$ to convert the stellar mass into total stellar 
luminosity, $L_*=M_*\times (L/M)$, and compare results with 
observations. We can see that model results match very well the 
observational data, including the nearly flat 
$\langle$[Fe/H]$\rangle$ value that is observed in ultra-faint 
dwarf galaxies. In Fig.~2 the colored boxes identify galaxies 
in different luminosity ranges, using the same color-code as in
Fig.~\ref{Fig:data}. These same colors will be used through the 
entire paper to distinguish among dwarf galaxies with different luminosities.

In our model, ultra-faint dwarf galaxies are predicted to be 
H$_2$-cooling minihaloes that form at $z > 7.5$, before 
photo-dissociation has suppressed gas-cooling and star formation in 
these small systems \citep{salvadori12}. In contrast, classical dSph 
galaxies are predicted to form at later times through the merging of 
smaller progenitors. On average we find that the more luminous is a 
galaxy, the more massive is predicted to be its host halo, and the 
lower its final assembly redshift \citep{salvadori09}. 
More luminous galaxies, therefore, keep accreting gas from the Galactic 
Medium while it is increasingly polluted with the heavy elements 
ejected by low-mass star-forming galaxies.\\

We finally underline that in our model we do not account for mass
transfer from binary companions. Thus, we can only investigate the 
incidence of CEMP(-no) stars that formed in carbon-enhanced environments. 
%%%%%%%%%%%%%%%%%%%%%%%%%%%%%%%%%%%%%%%%%%%%%%%%%%%%%%%%%%%%%%%%%%%%%%%%%%%%
\section{Results: the global picture}
\begin{figure*}
  \includegraphics[height=1.69\columnwidth]{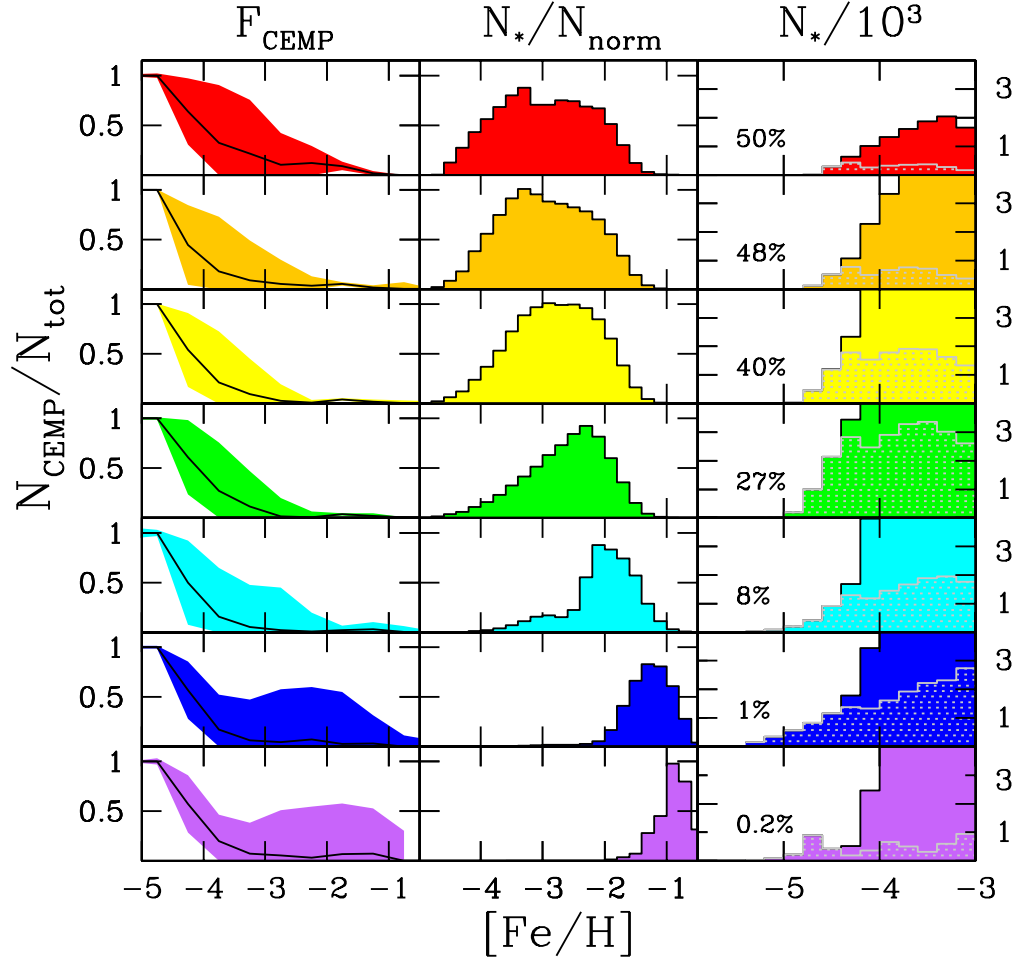} 
  \caption{Average properties of stars at different [Fe/H] in Milky Way 
    dwarf galaxies with increasing luminosity (from top to bottom):
    $L<10^4\Lsun$ (red), $10^4\Lsun<L<10^{4.5}\Lsun$ (orange), 
    $10^{4.5}\Lsun<L<10^5\Lsun$ (yellow), $10^5\Lsun<L<10^{5.5}\Lsun$ 
    (green), $10^{5.5}\Lsun<L<10^6\Lsun$ (cyan), $10^6\Lsun<L<10^7\Lsun$ 
    (blue), $10^7\Lsun<L<10^8\Lsun$ (bottom, violet).
    {\it Left panels}: average fraction of CEMP stars (solid line) with 
    $\pm1\sigma$ dispersion (shaded area). 
    {\it Middle panels}: average MDF normalized to the total number of stars 
    (histogram). {\it Right panels:} MDFs for stars at [Fe/H]$<-3$, showing 
    both the total number of stars (filled histograms) and the number of 
    CEMP-no stars (shaded gray histograms). The percentage shown is that 
    of stars at [Fe/H]$<-3$ with respect to the total.}
    \label{Fig:MDFs}
\end{figure*}

%%%%%%%%%%%%%%%%%%%%%%%%%%%%%%%%%%%%%%%%%%%%%%%%%%%%%%%%%%%%%%%%%%%%%%%%%%%%%%
We start by discussing the predictions of our cosmological model 
for the mean properties of dwarf galaxies in different luminosity 
ranges, which have been selected among all candidates in 50
different Milky Way realizations (Fig.~\ref{Fig:FeL}). The main 
results are summarized in Fig.~\ref{Fig:MDFs} where we show, for 
dwarf galaxies with increasing luminosities (from top to bottom), 
predictions for the average: i) fraction of CEMP stars $F_{\tt CEMP}$ 
(left), ii) normalized MDFs (middle), and iii) low-Fe tails of 
the MDFs (right). The fraction of CEMP stars with respect to the 
total, $F_{\tt CEMP}={N_{\tt L}}^{-1}\sum^{N_{\tt 
L}}_{i=1}N_*^{\tt CEMP}({\tt [Fe/H]})_i/N_*({\tt[Fe/H]})_i$, is
calculated by averaging among the total number of dwarf galaxies 
within a given luminosity range, $N_L$, where $N_*^{\tt CEMP}({\tt 
[Fe/H]})_i$ is the fraction of CEMP stars in the i-th galaxy, with 
a given [Fe/H] value. Similarly, we computed the mean MDFs by averaging among 
all $N_{\tt L}$ dwarf galaxies in a given $L$ range. For reference, 
Segue 1, Coma Berenice, and Ursa Major II have luminosities consistent 
with galaxies in the top panel (red), Bootes in the third (yellow), 
and Sculptor in the second from bottom (blue).
%%%%%%%%%%%%%%%%%%%%%%%%%%%%%%%%%%%%%%%%%%%%%%%%%%%%%%%%%%%%%%%%%%%%%%%%%%%%%%%%
%%%%%%%%%%%%%%%%%%%%%%%%%%%%%%%%%%%%%%%%%%%%%%%%%%%%%%%%%%%%%%%%%%%%%%

Several interesting features can be noted in Fig.~\ref{Fig:MDFs}:
(i) Independent of galaxy luminosity, we find that $F_{\tt CEMP}=1$ 
at [Fe/H]$\leq -5$, and it rapidly decreases towards higher [Fe/H], 
with a steeper decline in more luminous dwarf galaxies.
(ii) The shape of the normalized MDF dramatically changes with 
galaxy luminosity. In the least massive ultra-faint dwarf galaxies 
it is flat and covers a broad [Fe/H] range. As we move towards bigger 
galaxies, the MDF becomes more peaked, the peak is gradually
shifted towards higher [Fe/H] values, and the low Fe-tail turns 
into a smaller fraction of the total. 
(iii) Stars with [Fe/H]$<-3$ are predicted to be found in {\it all} 
dwarf galaxies, but their relative contribution to the overall 
stellar populations strongly decreases when the luminosity of the 
galaxy increases, as shown by labels in the panels. 
(iv) On average, the cumulative number of stars at [Fe/H]$<-4$ 
is roughly of the same order of magnitude, $\approx (1-2)\times 
10^3\Msun$, in all dwarf galaxies with the only exception of the 
faintest companions.

All these features are simply a result of the hierarchical galaxy 
formation process and the gradual {\it metal enrichment} and 
{\it reionization} of the environment out of which these galaxies 
form, which imprint their physical properties. In our scenario, the 
faintest dwarf galaxies are associated with H$_2$-cooling minihaloes, 
which virialize from the Milky Way environment at $7.5<z<12$,        
when $-5\leq$ [Fe/H]$_{GM}\leq-3$. The chemical enrichment proceeds 
very smoothly in these small systems, which transform gas into stars 
very inefficiently (Sec.~3, see also
\citealt{salvadori09,webster14,vincenzo14,romano15,blandhawthorn15}). 
Roughly a constant number of stars are formed at different evolutionary 
phases (or [Fe/H]), and the gas can reach high [Fe/H]  before 
being either completely evacuated by the cumulative effect of SNe 
explosions, or photo-heated by the increasing external ionizing radiation 
\citep{salvadori12}. This determines the characteristic shape of their 
MDF that is predicted, and observed, to be flat and to extend over a 
broad range of [Fe/H].

More luminous galaxies from via the assembly of these basics
building blocks and more massive progenitors, which form afterwards. 
In this bottom-up scenario all dwarf galaxies are expected to 
have, on average, some ultra-faint dwarfs among their parent haloes. 
Hence, they are predicted to share similar number of stars and MDF 
tails at the lowest [Fe/H]. However, the more luminous a galaxy, 
the less prominent is the contribution of the few thousand stars 
formed in these lowest-mass minihaloes into the final MDF. This 
is emphasized by the numbers reported in the right panels of Fig.~\ref{Fig:MDFs}. 
We can see that extremely metal-poor stars represent $\geq 50\%$
of the stellar population in ultra-faint dwarfs, while they are only
$\approx 0.3\%$ in dSph galaxies that have luminosity similar to Sculptor
or higher. For this reason, the low-Fe tails seem to gradually ``disappear'' 
in the normalized MDFs as the luminosity of the galaxies increases.
%%%%%%%%%%%%%%%%%%%%%%%%%%%%%%%%%%%%%%%%%%%%%%%%%%%%%%%%%%%%%%%%%%%%%%%%%%%%%%%
\begin{figure}
  \includegraphics[height=1.15\columnwidth]{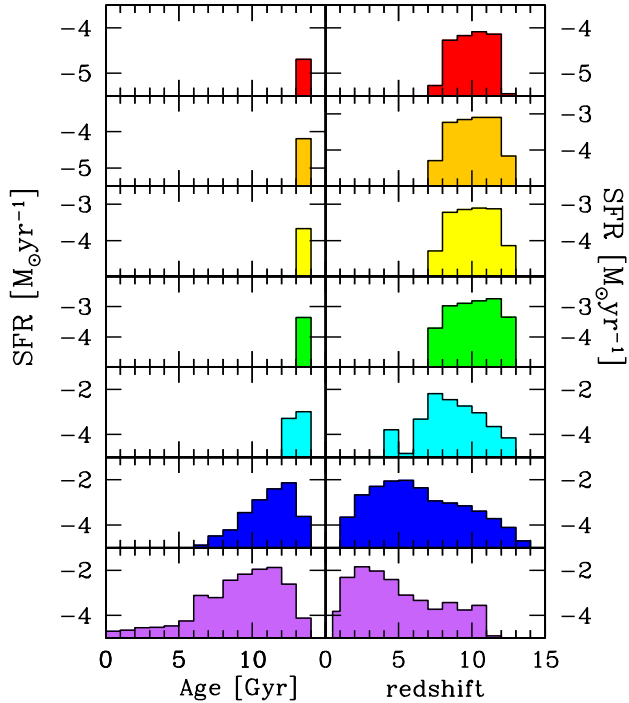}
  \caption{Predicted star formation rates of Milky Way dwarf galaxies 
     in different luminosity ranges (colors as described in Fig. 
     \ref{Fig:MDFs}) as a function of stellar ages ({\it left}) 
     and redshift ({\it right}).}
    \label{Fig:SFH}
\end{figure}
%%%%%%%%%%%%%%%%%%%%%%%%%%%%%%%%%%%%%%%%%%%%%%%%%%%%%%%%%%%%%%%%%%%%%%%%

Since more massive dwarf galaxies complete their assembly at lower 
redshifts, they can continue accreting gas from a 
Milky Way environment that is increasingly polluted by heavy elements. 
The average [Fe/H]$_{GM}$ value at the formation epoch of the main halo,
 where the bulk of the stars form, sets the lower limit of [Fe/H] in the
normalized MDF. 
Thus, on average, the more luminous a galaxy, the more pronounced 
is the shift of its MDF towards higher [Fe/H]. Furthermore, galaxies 
hosted by more massive dark matter haloes have bigger gas reservoir,
and they can more efficiently convert gas into stars because of their 
higher $T_{vir}>10^4$~K, which make them Ly$\alpha$ cooling haloes 
\cite[e.g.][]{maio07}. This causes a gradually more pronounced peak in their MDFs. 

The fraction of CEMP stars in dwarf galaxies with different luminosities 
is connected to all these previously presented effects. Looking at the 
right panels of Fig.~\ref{Fig:MDFs}, we can see that $N_{\tt CEMP}$ follows 
$N_*$ at [Fe/H]$<-4.5$, in all dwarf galaxies. These {\it CEMP-no}
stars are predicted to form in (progenitor) H$_2$-cooling minihaloes that 
have been predominantly polluted by primordial faint SNe. Because of the 
small amount of iron ($Y_{\tt Fe}\approx 4\times 10^{-7}$), and the huge 
amount of carbon ($Y_{\tt C}\approx 10^{-2}$) 
produced by faint SNe, long-lived Pop~II stars can already start to form 
in these small systems\footnote{Where the ``critical'' iron-abundance is 
settled by the yields of faint SNe: [Fe/H]$_{cr}=log({\tt Z_{cr}}/\Zsun)-
log(Y_{\tt Z}/Y_{\tt Fe})+log(M_{\tt Z}/M_{\tt Fe})_{\odot}$.} when the 
ISM is self-enriched up to [Fe/H]$_{cr}\approx -8$, which corresponds to 
[C/Fe]$\approx +4$, and $Z\approx Z_{cr}=10^{-4}\Zsun$. After the formation 
of Pop~II stars, normal SNII rapidly start to contribute to the chemical 
enrichment of both the ISM, gradually decreasing the [C/Fe] level while 
[Fe/H] rises, and the Milky Way environment, increasing $Z_{GM}$ up to 
$Z_{cr}$. In analogy to what is found in the Galactic halo \citep{caffau2011,
placco14} we predict that carbon-normal stars can start to form when 
[Fe/H]$\gsim -4.7$, which corresponds to $Z_{GM}>Z_{cr}$, and therefore 
to the disappearance of Pop~III stars \citep{debennassuti14}.

Thus, we predict that $F_{\tt CEMP}=1$ at [Fe/H]$\lsim-4.7$ in all dwarf 
galaxies, and that this fraction rapidly decreases with increasing [Fe/H], 
because of the larger contribution of SNII in both self-enriched and 
newly formed galaxies. CEMP-no stars populating the MDF at $-4<$
[Fe/H]$<-2$, predominantly form in environment polluted by primordial
faint SN and SNII. Thus they are all predicted to form in the very 
early stages of galaxy evolution. Interestingly, we can see that these 
CEMP-no stars efficiently form in dwarf galaxies like Bootes (yellow) 
and Draco (green), which are in the ``transition region'' between 
ultra-faint dwarfs and classical dSph galaxies (Fig.~\ref{Fig:FeL}). 
This is because these galaxies are associated to the most massive 
among minihaloes, with $T_{vir}\approx 10^4$~K, which 
host Pop~III stars as their least luminous companion, but have higher 
star formation efficiencies (Sec.~3). More massive, classical dSph 
galaxies, have higher probabilities to have among their progenitor 
haloes Pop~II galaxies that virialized from an environment pre-enriched 
up to [Fe/H]$_{GM}\geq -4$ by normal SNII \citep{salvadori07,debennassuti14}. 
Carbon-normal stars efficiently form in these galaxies, causing the 
steeper decline of $F_{\tt CEMP}$ with [Fe/H]. 

In stars with higher iron-abundance, [Fe/H]$> -2$, we find that the
carbon-enhancement might also come from low-metallicity ($Z\leq10^{-3}
\Zsun$) AGB stars, which evolve on longer time-scales. This causes 
the small increase of the $F_{\tt CEMP}$ values at [Fe/H]$\approx -1.5$, 
which is visible in the left panels of Fig.~\ref{Fig:MDFs}. Note that 
in our model we do not account for binary systems, so the carbon in these 
stars is not accreted from a companion, but it reflects the chemical 
composition of the birth environment. The products of AGB stars can 
be efficiently retained in the ISM of low mass galaxies after SNII 
have contributed to the chemical enrichment, and the galaxy can 
quietly evolve for a short period of time \cite[e.g.][]{salvadori12}. 
The production of s-process elements from $Z\leq 10^{-3}\Zsun$ AGB 
stars is still unclear \cite[e.g.][]{fisklock14} and we do not account 
for it in our work. However, these stars are most likely expected to 
be CEMP-s stars, although their properties are probably different from 
those formed in binary systems, which directly accrete material from 
an AGB companion.

Finally, we should emphasize that, although $F_{\tt CEMP}$ is steeper 
in more massive dwarfs, at [Fe/H]$<-3$ the results are all consistent 
within $\pm 1\sigma$ errors. As already discussed, this is because these 
different dwarf galaxies are expected to {\it share similar ancestors at 
high-redshift, namely low mass H$_2$-cooling minihaloes, which represent 
the birth environments of stars at [Fe/H]$<-3$}.

\subsection{Star formation histories}
Another way to consider these findings is to look at the predicted 
average star formation histories (SFHs) of dwarf galaxies, which 
are shown in Fig.~\ref{Fig:SFH} as a function of stellar ages, 
and formation redshift. We can see that all dwarf 
galaxies are predicted to host stars $>10$~Gyr old, which is 
what is observed in the Local Group \cite[e.g.][]{tolstoy09}. At very 
high redshifts, $z\geq 10$, we predict dwarf galaxies to share similar 
star formation rates, $\psi\approx 10^{-3}-10^{-4}\Msun/$yr, independent 
of their total luminosity. It is during these cosmic epochs that the low 
[Fe/H] tails of the MDFs are built up in the star-forming progenitor 
minihaloes.

Ultra-faint dwarf galaxies are predicted to stop forming stars prior
the end of reionization, $z>6$, because of heating by the external
UV background that prevents gas cooling, and ``sterilizes'' those systems   
that still have some leftover gas to fuel star formation \citep{salvadori12}.
This is consistent with observations of SFH in ultra-faint dwarfs
\cite[e.g.][]{brown14}. Note that some of these ``sterilized'' minihaloes
are predicted to evolve in isolation and survive until $z\approx 2$, when
they can be observed as very-metal-poor Damped Ly$\alpha$ systems
\citep{salvadori12}. These systems might re-start forming stars 
at later times \citep[e.g.][]{ricotti09,faerman13} and explain observations 
of HI-rich dwarf galaxies that show recent episodes of star formation (e.g. 
LeoT: \citealt{irwin07,dejong08,weisz12}).

More luminous galaxies continue to efficiently form stars to
much later times, and hence have more extended and complex SFHs. In the right
panels of Fig.~\ref{Fig:SFH} we see that $\psi$ is expected to gradually
increase towards lower redshifts, reaching the maximum at the main assembling
epoch of the host halo, and decreasing afterwards, when the galaxy is
assumed to evolve in isolation. The SFH that has been measured in the
Sculptor dSph galaxy \citep{deboer12}, is qualitatively consistent with
the one that we predict for dwarf galaxies with similar luminosity (blue):
it has a peak at $\approx 13$~Gyr, and then declines with cosmic time,
lacking stars $<6$~Gyr old.

In this global picture we can therefore explain the variation, with galaxy 
luminosity, of both the MDFs and star formation histories of dwarf galaxies.
%%%%%%%%%%%%%%%%%%%%%%%%%%%%%%%%%%%%%%%%%%%%%%%%%%%%%%%%%%%%%%%%%%%%%%%%%%%%%%%%
\section{Data comparison}
%%%%%%%%%%%%%%%%%%%%%%%%%%%%%%%%%%%%%%%%%%%%%%%%%%%%%%%%%%%%%%%%%%%%%%%%%%%%%%%%
We now test the results of our cosmological model against available 
observations for CEMP stars in nearby dwarf galaxies. We focus on 
three classes of dwarf galaxies that reside in different luminosity 
ranges, as shown in Fig.~\ref{Fig:FeL}: 
i) the ``classical'' dSph galaxy Sculptor, $L=10^{6.3\pm0.2}\Lsun$;
ii) the most luminous of the ultra-faint dwarf galaxies, Bootes, 
$L=10^{4.5\pm0.2}\Lsun$; and 
iii) the least luminous ultra-faint dwarfs, $L<10^4\Lsun$: the 
combination of SegueI, Coma Berenice, Ursa Major II, and LeoIV.   
As pointed out in Sec.~2, the available data for Bootes are 
mostly blind to CEMP sub-classes (s, r, and no). This represents 
a possible caveat for our comparison with CEMP-no models. Still, 
most of the CEMP stars found in this system have [Fe/H]$<-3$, which 
is the typical [Fe/H] range of CEMP-no stars \citep[e.g. Fig.~1 of
][]{norris13}.
 
Given the low number of [C/Fe] measurements in dwarf galaxies (see 
Fig.~\ref{Fig:Scl}a, \ref{Fig:dwarfs}a,d), we computed the uncertainty
of $F_{\tt CEMP}$ by using the results of \cite{Gehrels86} for Poisson 
statistics. We derived the 1$\sigma$ upper (lower) 
confidence limits as $F^{up}_{\tt CEMP}=1.841/N_*({\tt [Fe/H]})$ in case 
of non-detection ($F^{low}_{\tt CEMP}=0.0$), and as $F^{up}_{\tt CEMP}
=3.300/N_*({\tt [Fe/H]})$ for single detection ($F^{low}_{\tt CEMP}
=0.173/N_*({\tt [Fe/H]})$), where $N_*({\tt [Fe/H]})$ represents the 
total number of stars with available carbon measurements in different 
[Fe/H] ranges. So the fewer the measurements, the larger the uncertainties.
 
We used a Monte Carlo technique to randomly select from the mean theoretical 
MDFs, a number of stars equal to the total number of [Fe/H] measurements 
in different dwarf galaxies, $N_{tot}$. We then constructed the average 
$\pm1\sigma$ errors on the MDF by iterating this procedure 100 times. These 
errors, which are shown in Fig.~\ref{Fig:Scl}b and Fig.~\ref{Fig:dwarfs}b,c
together with the dispersion among different dwarf galaxies, provide an 
estimate of the data-driven uncertainties, i.e. they are the errors we 
always have to deal with while comparing theoretical models with small 
numbers of measurements.
%%%%%%%%%%%%%%%%%%%%%%%%%%%%%%%%%%%%%%%%%%%%%%%%%%%%%%%%%%%%%%%%%%%%%%%%%%%%%%
\subsection{Carbon-enhanced stars in Sculptor}
\begin{figure}
   \includegraphics[height=0.75\columnwidth]{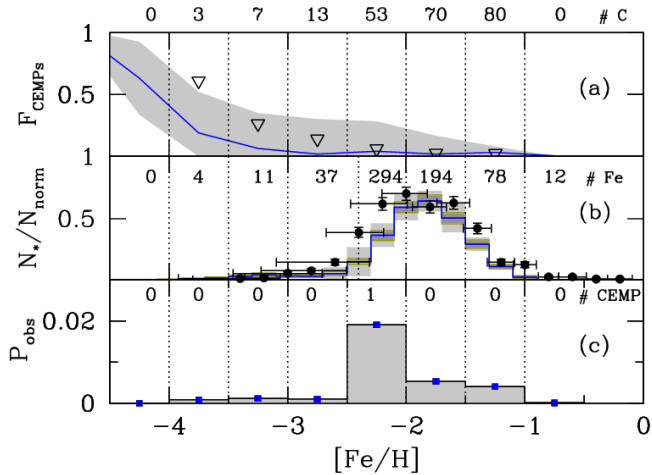}
    \caption{Predictions vs observations for the Sculptor dSph galaxy. 
     {\it Top panel:} the fraction of CEMP stars in different [Fe/H] bins. 
     The solid line shows the average value in the model, and the shaded 
     area the $\pm1 \sigma$ dispersion among different Sculptor-like 
     dwarf galaxies (as Fig.~\ref{Fig:MDFs}). Upside down triangles 
     are upper limits for the observed fraction of CEMP stars based on
     available data. Labels indicate the number of [C/Fe] measurements 
     in each [Fe/H] bin (references in Fig.~\ref{Fig:data}).
     {\it Middle panel:} the observed (points with errorbars) and predicted 
     MDF (histograms with shaded uncertainty). We show the $\pm1\sigma$ 
     dispersion among Sculptor-like dwarf galaxies in 50 Milky Way
     possible assembling histories (light gray), and among 100 Monte 
     Carlo sampling of the average MDF to the number of stars observed
     (dark gray). Labels indicate the
     number measurements in each [Fe/H] bin \citep{kirby11,starkenburg13}.
     {\it Bottom panel:} conditional probability to observe a star at 
     a given [Fe/H] that it is also carbon-enhanced. Labels indicate 
     the number of [C/Fe]$>0.7$ measurements in each [Fe/H].}
    \label{Fig:Scl}
\end{figure}

%%%%%%%%%%%%%%%%%%%%%%%%%%%%%%%%%%%%%%%%%%%%%%%%%%%%%%%%%%%%%%%%%%%%%%%%%%%%%%%%
The predicted average properties of Sculptor-like dwarf galaxies are 
shown in Fig.~\ref{Fig:Scl}. We note that there is only one detection 
of a CEMP-no star in Sculptor at $-2.5<$[Fe/H]$ \leq -2$ \citep{skuladottir15}, 
from which we get $F^{up}_{\tt CEMP}=0.06$. In the other [Fe/H] bins, 
with zero detections, $F^{up}_{\tt CEMP}$ is simply set by the number 
of [C/Fe] measurements. So the decline of $F^{up}_{\tt CEMP}$ towards 
higher [Fe/H] is simply a consequence of the larger number of stars 
observed at increasing [Fe/H].

From Fig.~\ref{Fig:Scl}a we can see that the theoretical predictions 
are consistent with $F^{up}_{\tt CEMP}$. As we already discussed in 
Sec.~4, because of the early chemical enrichment by primordial faint 
SN, $F_{\tt CEMP}$ rapidly declines with [Fe/H], ranging 
from unity at [Fe/H]$\leq-4.75$ down to values $<0.005$, at [Fe/H]$>-1$. 
Carbon-normal stars begin to form when [Fe/H]$\geq -4.7$, and in Sculptor 
they are predicted to be the majority of the stellar population ($>50\%$) 
already at [Fe/H]$\geq -4.25$. 

%%%%%%%%%%%%%%%%%%%%%%%%%%%%%%%%%%%%%%%%%%%%%%%%%%%%%%%%%%%%%%%%%%%%%%%%%
%%%%%%%%%%%%%%%%%%%%%%%%%%%%%%%%%%%%%%%%%%%%%%%%%%%%%%%%%%%%%%%%%%%%%%%%%%%%%
This result has two implications: i) to catch second-generation stars 
imprinted {\it mainly} by primordial faint SNe in Sculptor, we should 
look among stars at [Fe/H]$<-4.75$, as in the Galactic halo; ii) to 
increase the probability of finding CEMP stars in Sculptor we should
follow up [Fe/H]$<-3$ stars, for which we predict $F_{\tt CEMP}\geq 0.065$.
Our ability to find these CEMP stars will remain naturally limited by 
the absolute number of stars that exist at these low [Fe/H]. Our model
predicts, and this is supported by observations, that stars at [Fe/H]$<-3$ 
are {\it intrinsically} rare in luminous, Sculptor-like galaxies, only 
representing $< 3\%$ of the total stellar population (see Fig. 
\ref{Fig:MDFs}). This point is highlighted in Fig.~\ref{Fig:Scl}b, 
where we compare the predicted and observed MDFs, that are normalized 
to the total number of stars. Noticeably, stars at [Fe/H]$<-3$ are 
almost invisible in the normalized function, which is dominated by 
more Fe-rich stars. Clearly, the paucity of stars at this low [Fe/H] 
makes it very challenging to search for CEMP-no stars in Sculptor, 
where only the brighter stars at the tip of the RGB can be followed up. 
Between $-3.5\leq$[Fe/H]$<-3$, for example, we should roughly double 
the number of stars {\it found} in Sculptor to be able to observe one 
CEMP-no star among them. 

Thus, we can ask a different question: what is the joint probability 
to observe a star that has a given [Fe/H] value, and that is also 
carbon-enhanced? Fig.~\ref{Fig:Scl}c shows this joint probability 
distribution. This has been computed by combining the two independent 
functions: the fraction of CEMP stars {\it predicted} by the model, 
and the {\it observed} MDF normalized to the total number of stars, 
$P_{obs}=F_{\tt CEMP}\times N_*/N_{tot}$. The probability, $P_{obs}$, 
is maximal at [Fe/H]$\approx -2$, where we predict the highest absolute 
number of CEMP stars. {\it This result naturally explains why the 
first carbon-enhanced star in the Sculptor dSphs has been serendipitously 
found at such a high [Fe/H]}. In conclusion, the observed MDF, the 
derived upper limits for $F_{\tt CEMP}$, and the iron-abundance of 
the only CEMP star detected in Sculptor, are all consistent with the 
hypothesis that faint SNe were the main contributors to primordial 
metal enrichment.
%%%%%%%%%%%%%%%%%%%%%%%%%%%%%%%%%%%%%%%%%%%%%%%%%%%%%%%%%%%%%%%%%%%%%%%%%
\subsection{Carbon-enhanced stars in ultra-faint dwarfs}
\begin{figure*}
  \centerline{\psfig{figure=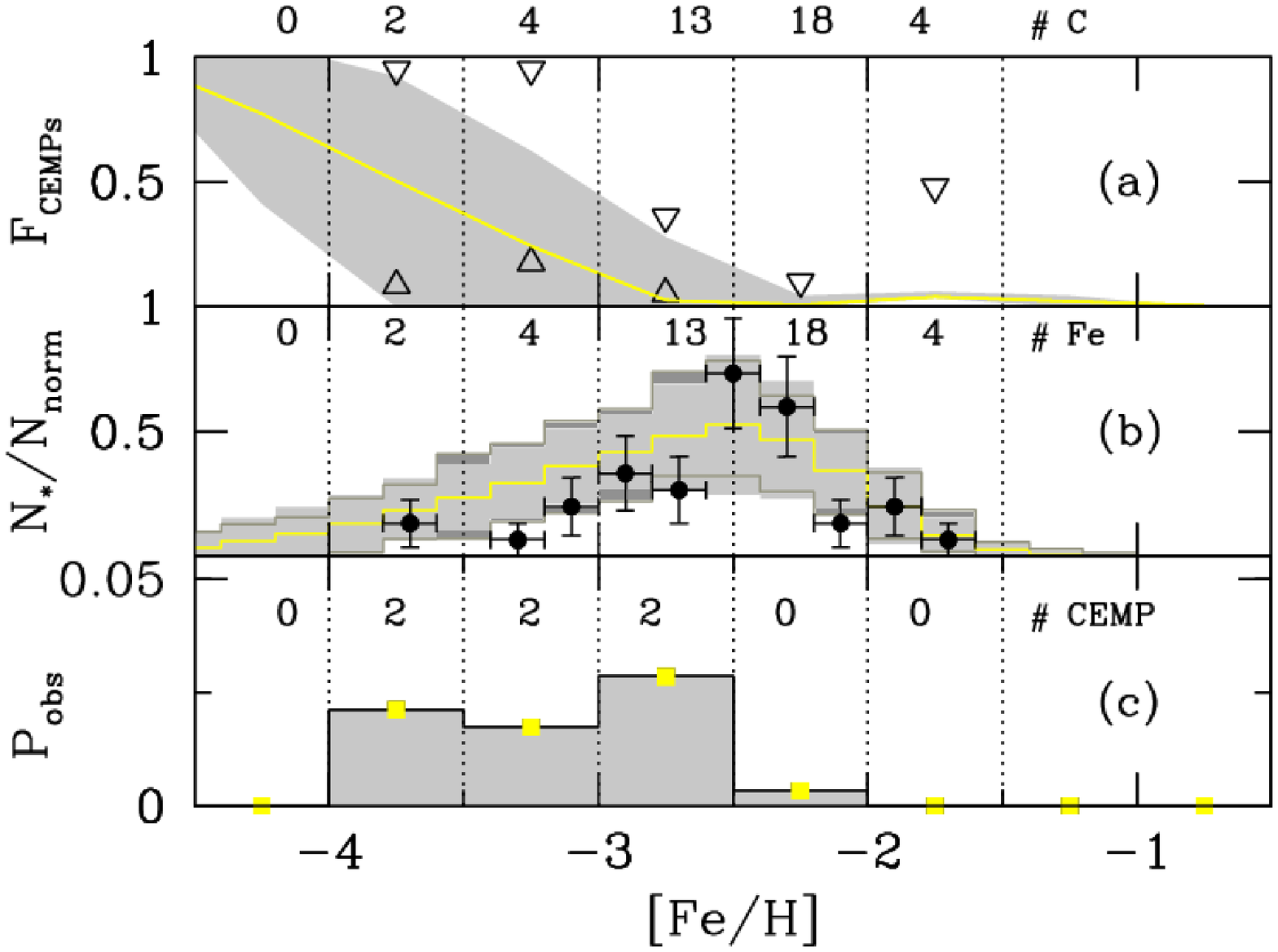,height=6.55cm}
  \psfig{figure=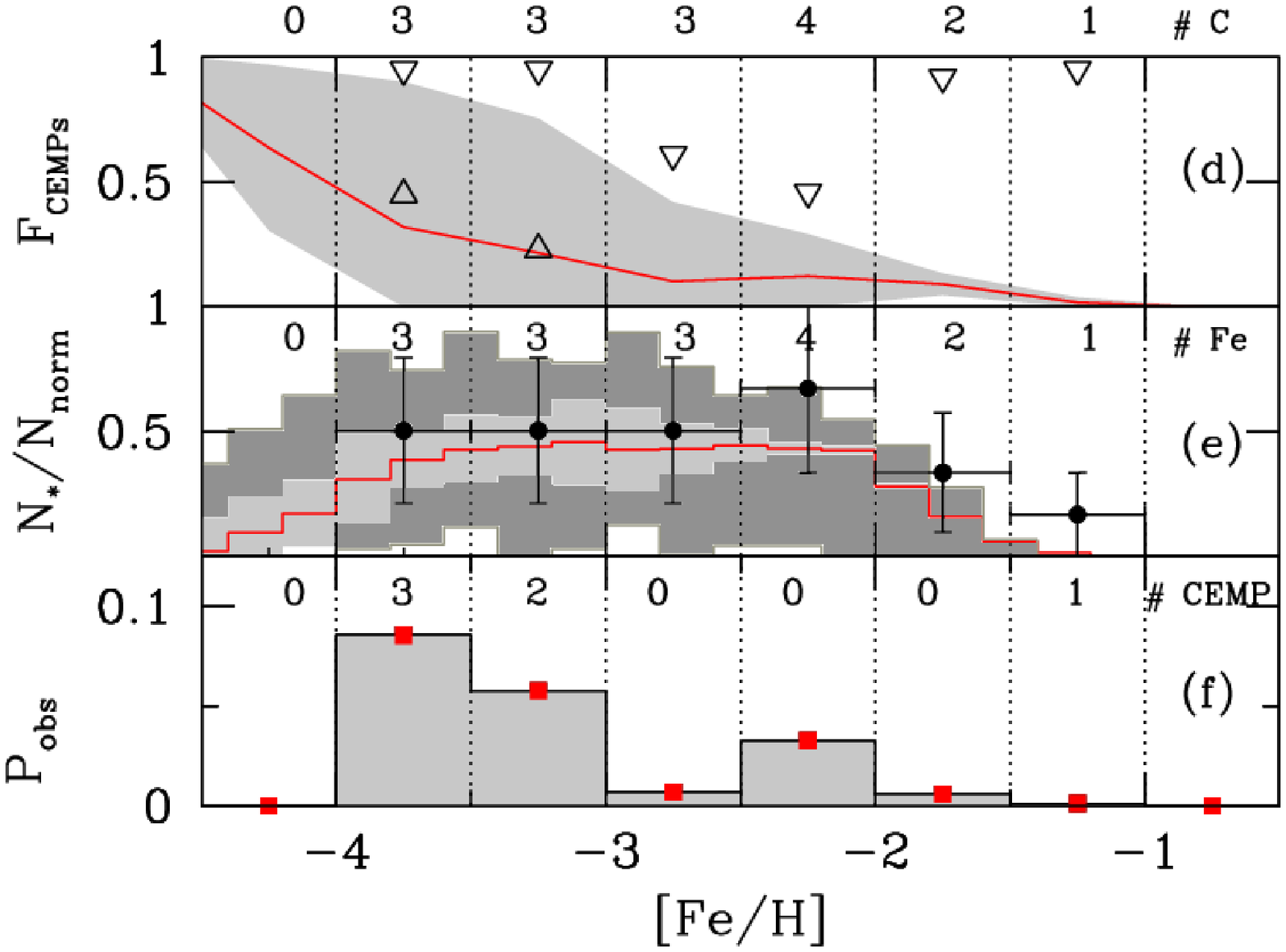,height=6.55cm}}
    \caption{Same as Fig.~\ref{Fig:Scl} but for Bootes-like dwarf galaxies
    ({\it left}), and for the least luminous ultra-faint dwarfs with 
    $L<10^4\Lsun$ ({\it right}).  
    Measurements are the same reported in Fig.~\ref{Fig:data}. 
    For $F_{\tt CEMP}$ we show both upper and lower limits from
    observations (see text).}
    \label{Fig:dwarfs}
\end{figure*}

%%%%%%%%%%%%%%%%%%%%%%%%%%%%%%%%%%%%%%%%%%%%%%%%%%%%%%%%%%%%%%%%%%%%%%%%%
We now explore the predictions of our model for less luminous 
ultra-faint dwarf galaxies, where many CEMP stars have been found 
(Fig.~\ref{Fig:data}). In Fig.~\ref{Fig:dwarfs} we show the results 
for Bootes-like dwarf galaxies and for the least luminous ultra-faint 
dwarfs with $L<10^4\Lsun$. As for Sculptor we select these galaxies 
on the bases of their total luminosity among all the available 
candidates predicted by the model (Fig.~\ref{Fig:FeL}). 

By inspecting Fig.~\ref{Fig:dwarfs}a and Fig.~\ref{Fig:dwarfs}d, 
we can see that because of the low total number of stars observed 
in these small systems ($\approx 40$ in Bootes, and $\approx 15$ in 
the faintest ultra-faint dwarf galaxies combined all together) the 
upper/lower limits on $F_{\tt CEMPs}$ are extremely high/low, and 
hence naturally consistent with model predictions. In these ancient 
systems, the fraction of RGB stars with respect to the total is 
estimated\footnote{We used PARSEC isochrones \citep{bressan12,chen14,
tang14}, and the CMD generator available at http://stev.oapd.inaf.it/cmd} 
to be $N_{\tt RGB}/N^{tot}_*\approx 0.001$. This implies that in both
cases $\geq 50\%$ of RGB stars have been already followed-up\footnote{For 
the ultra-faint dwarfs this only is true if we consider the {\it combination} 
of the four systems at $L<10^4\Lsun$. In Segue~I all RGB
stars have already been followed up.}

We note that in the model, the dispersion among different galaxies is 
huge in the case of the faintest dwarf galaxies (Fig.~\ref{Fig:dwarfs}d), 
and at [Fe/H]$=-3.5$ the $+1\sigma$ error is consistent with $F_{\tt CEMP}
\approx 0.8$. This is because these faint galaxies are associated 
with the least massive H$_2$-cooling minihaloes that are able to form 
stars, and which evolve in isolation. These systems can either virialize 
from a primordial birth-environment, $Z<Z_{cr}$, and hence $F_{\tt CEMP}
\approx 1$ for [Fe/H]$< -3$ because they host primordial faint SN, or 
from a medium that has been pre-enriched up to $Z>Z_{cr}$ by the products 
of SNII. In the latter case they will only host Pop~II stars, and hence
$F_{\tt CEMP}\approx 0$. These different formation paths cause the large 
spread in $F_{\tt CEMP}$ at [Fe/H]$<-3$. According to our model, the large 
fraction of CEMP-no stars observed in the faintest dwarfs at [Fe/H]$<-3$ 
($\approx 80\%$) suggests that these systems are {\it truly Pop~III galaxies}, 
which have experienced primordial star formation. However, we clearly 
need better measurement statistics. A quest for more data also emerges 
while comparing the predicted and observed MDFs (Fig.~\ref{Fig:dwarfs}e). 
In the faintest dwarf galaxies, the errors induced by the low number of
measurements (15 stars) are larger than the dispersion of the model among 
different dwarf galaxies and Milky Way merger histories. 
In contrast, these errors are equal (lower) than the intrinsic model 
uncertainties in Bootes (Sculptor), where $\approx 40$ stars 
($\approx 700$) have been observed.
%%%%%%%%%%%%%%%%%%%%%%%%%%%%%%%%%%%%%%%%%%%%%%%%%%%%%%%%%%%%%%%%%%%%%%%%%
%%%%%%%%%%%%%%%%%%%%%%%%%%%%%%%%%%%%%%%%%%%%%%%%%%%%%%%%%%%%%%%%%%%%%

Fig.~\ref{Fig:dwarfs}c, f show the joint probability to observe a star 
in a given [Fe/H] range, that is also carbon-enhanced. If we compare 
this function with the one derived for Sculptor (Fig.~\ref{Fig:Scl}c), 
we note a clear trend: the fainter the galaxy is, the higher is the
{\it overall} probability to observe a star that is also carbon-enhanced, 
which is the integral of $P_{obs}$ over the [Fe/H] range. This 
explains why it is much easier to find carbon-enhanced stars in these 
small systems. Moreover, $P_{obs}$ is maximum at lower [Fe/H] in less 
luminous dwarf galaxies. For Bootes we find that $P_{obs}$ is maximum 
at $-4 \leq$[Fe/H]$\leq -2.5$, while for the faintest dwarfs at 
[Fe/H]$\leq -3$. In both cases the predicted [Fe/H] ranges coincide 
with [Fe/H] values of observed CEMP-no stars in these small systems.

Finally, we should note that for the faintest dwarf galaxies the model 
predicts a non-negligible fraction of CEMP stars at [Fe/H]$>-2$ (upper 
panel). As explained in Sec.~4, these stars are predicted to form in 
environment polluted by the products of AGB stars with $Z<10^{-3}\Zsun$, 
which can be retained by these small galaxies when SN have already exploded. 
Interestingly, a CEMP-s star at [Fe/H]$\approx -1.5$ that does not 
show evidence for a binary companion has been recently detected in 
Segue 1 \citep{frebel14}. So, although more data are required to solidly 
assess if this star is in a binary system, this observation supports the 
idea that the physical mechanism we propose may contribute to the formation 
of CEMP-s in these small systems.
%%%%%%%%%%%%%%%%%%%%%%%%%%%%%%%%%%%%%%%%%%%%%%%%%%%%%%%%%%%%%%%%%%%%%%%%%%%%%
\section{The low-[Fe/H] tail of Sculptor}
%%%%%%%%%%%%%%%%%%%%%%%%%%%%%%%%%%%%%%%%%%%%%%%%%%%%%%%%%%%%%%%%%%%%%%%%%%%%%%
The results of our cosmological model show that CEMP-no stars, similar
to those found in the Galactic halo and ultra-faint dwarfs, are not 
necessarily missing in classical dSph galaxies, such as Sculptor, but 
they could be {\it hidden} among rare [Fe/H]$<-3$ stars, which represent 
$<3\%$ of the overall stellar population. The possible existence
of a CEMP-no star at [Fe/H]$=-3.77$ (see Section 2) might be the first 
confirmation that this is indeed the case. Sculptor-like dSph galaxies, 
furthermore, are predicted to be among the best systems to find the most 
iron-poor stars, because their MDFs might extend down to [Fe/H]$<-5$, if 
faint SN dominated the early metal enrichment Fig.~\ref{Fig:MDFs}.
The key question is: how much do we need to enlarge the current stellar 
sample to unveil the lowest-Fe tail of the Sculptor MDF and catch the most 
pristine CEMP stars?

\begin{figure}
  \includegraphics[height=0.99\columnwidth]{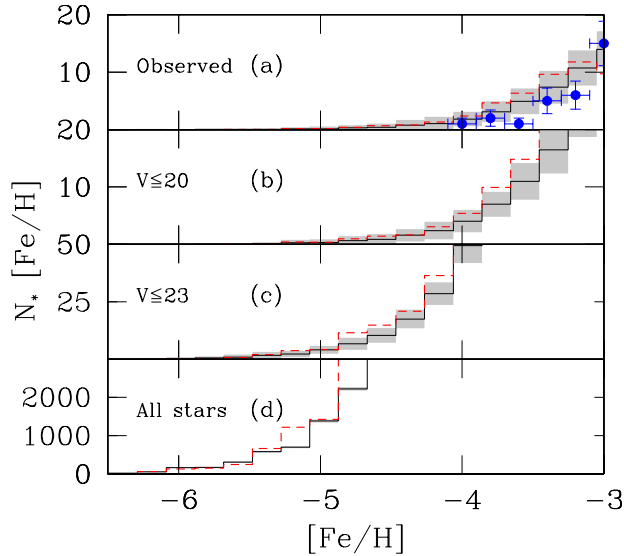}
  \caption{Number of stars at [Fe/H]$<-3$ that can be observed in
     Sculptor by increasing the sample of [Fe/H] measurements.
     From top to bottom we show results for:
     i) the current statistics ($\approx 700$ RGB stars), where the 
     available data are shown as points with Poissonian errorbars 
     (Fig.~5 for references); ii) stars with $V\leq20$; iii) stars 
     with $V\leq 23$; iv) all stars. Shaded area show the $\pm 
     1\sigma$ errors obtained from the Monte Carlo random selection 
     technique (see text). Solid histograms with shaded area 
     are results for our fiducial model, $Z_{cr}=10^{-4.15}\Zsun$. 
     Red dashed histograms show the same results for the case 
     $Z_{cr}=10^{-6}\Zsun$.}
    \label{Fig:low_tail}
\end{figure}

Fig.~\ref{Fig:low_tail} shows how many stars at [Fe/H]$< -3$ are 
predicted to emerge in Sculptor-like dwarf galaxies by increasing 
the total number of [Fe/H] measurements. Fig.~\ref{Fig:low_tail}a
exhibits results for the actual number of observed stars, $\approx 700$ 
stars, which roughly correspond to $\approx 25\%$ of all RGB stars 
in Sculptor. Model predictions agree quite well with observations.
Fig.~\ref{Fig:low_tail}b shows how the low-Fe tail of Sculptor 
will appear by following up all stars down to magnitudes $V\leq 20$, 
which corresponds to $\approx 1600$ stars in the observed Color 
Magnitude Diagram \citep[][]{deboer12}. We find that in this case
we might be able to observe $12\pm 8$ stars with [Fe/H]$\leq -4$, 
the $(40\pm 20)\%$ of which are predicted to be CEMP-no stars. 
These observations can be achieved with present-day instrumentation 
(e.g. ESO VLT with X-Shooter, FLAMES, and UVES, or Keck/DIMOS) to 
add crucial information to our understanding of the properties of 
first stars and early galaxy formation processes. 
See for example \cite{kirby09}, who used Keck/DIMOS to obtain 
medium-resolution spectra ($6400-9000$ \AA, $R\approx 6500$) for
$V\approx 20$ stars in the central region of Sculptor and derive their
C and Fe abundances. Future generation of telescopes will allow us 
to measure [Fe/H] for stars below the main sequence 
turn-off in nearby galaxies, like Sculptor. For example, the limit 
given for MOSAIC spectroscopy on the ESO-ELT (Evans et al. 2015) 
is $V\approx 25$ at $R\approx 15,000-20,000$. This will dramatically 
increase the number of stars that can be observed at high spectral 
resolution in nearby galaxies. The total number of RGB and MS stars 
in Sculptor with $V\leq 23$ is $\approx20000$, and from these we can 
expect $80\pm 22$ stars at [Fe/H]$\leq -4$ (Fig.~\ref{Fig:low_tail}c), 
and $16\pm 10$ stars with [Fe/H]$\leq-4.7$, where the fraction of CEMP-no 
stars is expected to be $100\%$.
With these observations, furthermore, we might be able to accurately 
constrain the critical metallicity value. As already shown in 
\cite{salvadori07} (Fig.~7), decreasing $Z_{cr}$ in the most feasible 
range, $Z_{cr}\approx 10^{-4}-10^{-6}\Zsun$, mainly affects the number 
of long-lived stars at [Fe/H]$<-4$. Thus, the differences between $Z_{cr}$ 
models become observable in Sculptor when a significant number of stars 
are followed-up, and the lowest-Fe tail starts to emerge (Fig.~7c).
%%%%%%%%%%%%%%%%%%%%%%%%%%%%%%%%%%%%%%%%%%%%%%%%%%%%%%%%%%%%%%%%%%%%%%%%%%%%%
\section{Summary and discussion}
%%%%%%%%%%%%%%%%%%%%%%%%%%%%%%%%%%%%%%%%%%%%%%%%%%%%%%%%%%%%%%%%%%%%%%%%%%%%%%
%%%%%%%%%%%%%%%%%%%%%%%%%%%%%%%%%%%%%%%%%%%%%%%%%%%%%%%%%%%%%%%%%%%%%%%%
We used a robust, data-constrained merger-tree model for 
the possible formation histories of the Milky Way and its 
dwarf satellites to predict the frequency of CEMP-no 
stars in nearby dwarf galaxies \citep[e.g.][]{salvadori08,
salvadori09,salvadori12,salvadori14}. We have shown that 
the model can successfully explain the variation of the 
{\it average} MDFs and SFHs observed in dwarf galaxies 
with increasing luminosities \citep[e.g.][]{kirby11,
weisz14} by accounting for star-formation in H$_2$-cooling 
mini-haloes, $M=10^{6.5}-10^8\Msun$, with a self-consistent 
treatment of the reionization and metal enrichment of the 
Milky Way environment.
 
By assuming that primordial faint SN, with mixing and fallback, 
dominated the early chemical enrichment, as suggested by 
observations of Galactic halo stars \cite[e.g.][]{iwamoto05,
cooke14,marassi14,debennassuti14}, we analyze the {\it average} 
properties of dwarf galaxies in different luminosity ranges, 
and show that:
\begin{itemize} 
\item CEMP-no stars should exist in {\it all} dwarf galaxies within 
the observed luminosity range, $10^{2.5}\Lsun< L< 10^{7.5}\Lsun$.
\item Independent of galaxy luminosity the relative 
fraction of CEMP-no stars increases towards lower [Fe/H], 
reaching $100\%$ at [Fe/H]$\lsim-4.7$. 
\item As the galaxy luminosity increases, the overall 
probability to observe CEMP-no stars decreases, and   
the [Fe/H] range in which they are most likely to be 
found is shifted towards higher values.
\item In classical Sculptor-like dSph galaxies, the 
probability to find CEMP-no stars is almost an 
order of magnitude lower than in the faintest Milky 
Way companions, $L<10^4\Lsun$, and it is maximal, 
$P_{obs}=0.02$ at [Fe/H]$\approx -2$. 
\end{itemize}
The results explain why it is easier to discover
CEMP-no stars in ultra-faint dwarf galaxies than 
in more luminous classical dSphs \citep[e.g.][]{norris10,
lai11,frebel14}, and also why the only CEMP-no star observed 
in Sculptor has been found at an unexpectedly high 
[Fe/H] \citep{skuladottir15}. In particular, our model 
shows that CEMP-no stars at [Fe/H]$\lsim -4.7$ form in 
environments predominantly polluted by primordial faint 
SN, which therefore have extremely high [C/Fe] values.
On the other hand, those at $-4.7\lsim$[Fe/H]$\lsim-2.0$ are 
imprinted by both primordial faint SN and low metallicity, 
$Z<10^{-3}\Zsun$, SNII, in agreement with recent findings
by Bonifacio et al. (2015). At [Fe/H]$\gsim -2.0$, we find 
that CEMP stars can also form in an ISM polluted by the 
products of AGB stars with $Z<10^{-3}\Zsun$. Hence, 
they may potentially be enriched by s-process neutron 
capture elements (i.e. CEMP-s), which are not included
in our model.

Our findings for CEMP-no stars, are a consequence of both 
the extremely low Fe-production, and high C, from faint SN, 
and of the cosmological context in which the hierarchical 
assembly of dwarf galaxies occurs.
H$_2$-cooling minihaloes, or ultra-faint dwarf galaxies, 
$L<10^5\Lsun$, are thus predicted to be the high-redshift 
progenitors of more massive ``classical'' dSph galaxies, 
and the environment of formation for stars with [Fe/H]$<-3$. 
The low-Fe tails of dwarf galaxies are thus expected to 
form in these common building blocks, some of which 
experienced Pop~III star formation \citep{salvadori14}. 
Such low-Fe tails can extend down to [Fe/H]$<-4.7$ if 
built-up upon the chemical products of primordial faint SN.

As the galaxy luminosity increases our model shows that 
the average MDFs become more {\it peaked}, and shifted 
towards {\it higher} [Fe/H] values, as is observed.
Thus, stars at [Fe/H]$<-3$ become a {\it lower fraction} 
of the total stellar populations: from $\geq 40\%$
for ultra-faint dwarf galaxies, $L<10^5\Lsun$, to $<0.2\%$ 
for $L>10^7\Lsun$. This is why in more massive dwarf 
galaxies the overall probability to observe a CEMP-no 
star decreases, and the [Fe/H] range in which they are 
more likely to be found increases.
In Sculptor, $L\approx 10^{6.3}\Lsun$, we find that the 
fraction of CEMP-no stars monotonically decreases from 
$F_{\tt CEMP}\approx 1$ at [Fe/H]$\leq -4.75$ down to 
$F_{\tt CEMP}<0.005$ at [Fe/H]$\geq -1$. However, the 
probability to observe one of these stars is higher at
$-2.5<$[Fe/H]$\leq -2$ because more stars can be observed 
in this range. 

A key prediction of our work is that galaxies with different 
luminosity are expected to share, on average, similar MDF 
tails {\it and} fractions of CEMP-no stars at the lowest [Fe/H]. 
At the moment, this hypotheses is supported by the observations
of a very similar fraction of CEMP stars at [Fe/H]$<-3$ in the 
Galactic halo and in the {\it overall} sample of nearby 
dwarf galaxies, $F_{\tt CEMP}(\leq -3)\approx (43-42)\%$. However, 
more data need to be collected for the Milky Way companions.
This also emerges from our analysis of the model uncertainties,
which in the faintest dwarf galaxies are dominated by the 
low number statistics.

Finally, we should be clear that our model cannot exclude
alternative scenarios for the formation of CEMP-no stars, 
such as metal pollution by massive rotating primordial stars 
\citep[e.g.][]{meynet06,meynet15}.
Furthermore, we did not explore the effects of different
primordial Initial Mass Functions on the fraction of CEMP-no
stars in dwarf galaxies. Instead, we simply worked under the 
hypothesis that Pop~III stars evolve as faint SN with mixing 
and fallback, since these pristine stars should dominate the 
early chemical enrichment to successfully explain observations 
of CEMP-no stars in the Galactic halo \citep[e.g.][]{cooke14,
debennassuti14}. Our findings show that to further test the 
predominant role of primordial faint SN in the early Universe
we need to at least double the current stellar sample in 
Sculptor, by measuring [Fe/H] and [C/Fe] in stars in different 
region of 
the galaxy down to magnitudes $V\leq 20$. Deeper observations 
($V\leq 23$) should enable the discovery of peculiar CEMP-no 
stars in Sculptor similar to those found in the Galactic halo 
at [Fe/H]$<-4$. This will provide crucial information not only 
on the nature of first stars and critical metallicity value, 
but also on the formation of the Galactic halo, and hence on the 
underlying hierarchical galaxy formation process.
%%%%%%%%%%%%%%%%%%%%%%%%%%%%%%%%%%%%%%%%%%%%%%%%%%%%%%%%%%%%%%%%%%%%%%%%%%%%%%%
\section*{Acknowledgements}
We thank the anonymous referee for his/her useful insights, and
very constructive report. S.~Salvadori is grateful to P. Bonifacio 
and M. de Bennassuti for useful discussions. She warmly thanks J. 
Norris and E. Kirby for careful reading the first version of the 
paper, and C. Chiappini for key insights on early results. Finally, 
she is grateful to the Netherlands Organization for Scientific Research,
which supports her research through a VENI grant 639.041.233.
%%%%%%%%%%%%%%%%%%%%%%%%%%%%%%%%%%%%%%%%%%%%%%%%%%%%%%%%%%%%%%%%%%%%%%%%%%%%%%%%
\bibliographystyle{mn2e}
%\bibliography{salvadori}

\label{lastpage} 
\end{document}